\documentstyle[aps,prl,preprint,floats,epsfig]{revtex}

\begin{document}

\preprint{\tighten\vbox{\hbox{\hfil CLNS 00/1698}
                        \hbox{\hfil CLEO 00-22}
}}

\title{Observation of New States Decaying into $\Lambda_c^+\pi^-\pi^+$}

\author{CLEO Collaboration}
\date{\today}

\maketitle
\tighten

\begin{abstract} 
Using $13.7 fb^{-1}$ of data recorded by the 
CLEO detector
at CESR, we investigate the spectrum of charmed baryons 
which decay into $\Lambda_c^+\pi^-\pi^+$
and are more massive than
the $\Lambda_{c1}$ baryons. 
We find evidence for two new states: one 
is broad and has an invariant mass roughly 480 MeV above 
that of the $\Lambda_c^+$ baryon; the 
other is narrow with an invariant mass of 
$596\pm 1\pm 2\ $ MeV
above the $\Lambda_c^+$ mass. 

\end{abstract}
\newpage

{
\renewcommand{\thefootnote}{\fnsymbol{footnote}}

\begin{center}
M.~Artuso,$^{1}$ R.~Ayad,$^{1}$ C.~Boulahouache,$^{1}$
K.~Bukin,$^{1}$ E.~Dambasuren,$^{1}$ S.~Karamov,$^{1}$
G.~Majumder,$^{1}$ G.~C.~Moneti,$^{1}$ R.~Mountain,$^{1}$
S.~Schuh,$^{1}$
% ...
\end{center}
 
\small
\begin{center}
$^{1}${Syracuse University, Syracuse, New York 13244}\\
% ...
\end{center}

\setcounter{footnote}{0}
}
\newpage

Studies in the last decade have revealed a rich spectroscopy of charmed baryon states.
Baryons consisting of a charmed quark and two light (up or down) quarks are denoted the
$\Lambda_c$ and $\Sigma_c$ baryons, depending on the symmetry properties of the wave function.
All three of the ground state $J^P$$=$${1\over{2}}^{+}$ $\Sigma_c$
and all three of the ground state $J^P$$=$${3\over{2}}^{+}$ $\Sigma_c^*$ 
particles have been identified. 
Knowledge
of orbitally excited states in the sequence is presently
limited to the observation of 
two states
decaying into $\Lambda_c^+\pi^+\pi^-$\cite{LCS}. These have been identified as the 
$J^P$$=$${1\over{2}}^-,
{3\over{2}}^-$ $\Lambda_{c1}^+$ particles, where the numerical subscript denotes one 
unit of light quark angular momentum. 
There must be many more excited states still to be found. 
Here we detail the results of 
a search for such states that decay into a $\Lambda_c^+$ baryon with the emission of
two oppositely charged pions.

The data presented here 
were taken using the CLEO II and CLEO II.V detector configurations
operating at the Cornell 
Electron Storage Ring.
The sample used in this analysis corresponds to
an integrated luminosity of 13.7 $fb^{-1}$ from data
taken on the $\Upsilon(4S)$ 
resonance and in the continuum at energies just 
below the $\Upsilon(4S)$.
Of this data, 4.7 $fb^{-1}$ was taken with the CLEO II detector\cite{KUB},
in which we detected charged tracks using a cylindrical drift chamber system inside
a solenoidal magnet and photons using an electromagnetic
calorimeter consisting of 7800 CsI crystals.
The remainder of the data was taken with the CLEO II.V 
configuration\cite{HILL}, which has
upgraded charged particle measurement capabilities, but the same
CsI array to observe photons.

In order to obtain large statistics we reconstructed the $\Lambda_c^+$
baryons using 15 different decay modes.
\footnote{Charge conjugate modes are implicit throughout.} 
Measurements of the branching
fractions into these modes have previously been presented by the CLEO
collaboration\cite{LAMC}, and the general procedures for finding
those decay modes can be found in these references.
For this search and data set, the exact analysis used has been optimized
for high efficiency and low background.
Briefly, particle identification of $p,K$, and $\pi$ candidates was performed
using specific ionization measurements in the drift chamber
and, when available, time-of-flight measurements. Hyperons were found by
detecting their decay points separated from the main event vertex.

We reduce the combinatorial background, which is highest for
charmed baryon candidates with low momentum, by applying a cut on the
scaled momentum $x_p=p/p_{max}$.
Here $p$ is the momentum
of the charmed baryon candidate, $p_{max}=\sqrt{E^2_{bm}-M^2}$,
$E_{bm}$ is the beam energy, and
$M$ is the invariant mass of the candidate. 
Note that charmed baryons produced from decays of $B$ mesons are
kinematically limited to $x_p < 0.4$.
Requiring $x_p > 0.5$,
we fit the invariant mass distributions for these modes to a sum
of a Gaussian signal and a low-order polynomial background.
Combinations within $1.6 \sigma$ of the mass of the
$\Lambda_c^+$ in each decay mode are taken as $\Lambda_c^+$ candidates,
where the resolution, $\sigma$,  of each decay mode is taken from a
GEANT-based\cite{GEANT} Monte Carlo simulation
for the two detector configurations separately. In this $x_p$ region,
we find a total yield of $\Lambda_c^+$ signal
combinations of $\approx$ 58,000, and a signal to background ratio
$\approx $$5$$:$$6$.
This is the same sample of $\Lambda_c^+$ baryons that has been used in our 
discovery of 
the $\Sigma_c^{*+}$\cite{sigmastarplus}.
This $x_p$ restriction
was released before continuing with the analysis as we prefer to apply
such a criterion only on the parent $\Lambda_c^+\pi^+\pi^-$ combinations.

The $\Lambda_c^+$ candidates were then
combined with two oppositely charged $\pi$ candidates in the event. To obtain 
the best resolution, the trajectories of the $\pi$ candidates were constrained
to pass through the main event vertex. 
The large combinatoric backgrounds and the hardness of the momentum
spectrum of the known excited charmed baryons led us to place
a cut of $x_p > 0.7$ on the combination. 
Figure 1 shows the mass difference spectrum,
$\Delta M_{\pi\pi} = M(\Lambda_c^+\pi^+\pi^-)-M(\Lambda_c^+)$, for
the region above the well-known $\Lambda_{c1}$ resonances.
Also shown in Figure 1 are combinations formed using appropriately scaled
sidebands of 
the $\Lambda_{c}^{+}$ signal.
An attempt to fit the upper plot in Figure 1 to only a second order polynomial shape yields
an unacceptable $\chi^2$ of 184 for 77 degrees of freedom.
However, if it is fit to the sum of a second order polynomial and two
Gaussian signals, the resultant $\chi^2$ is 59 for 71 degrees of freedom. 
Of these two signals, the lower one has a yield of 
$997^{+141}_{-129}$, $\Delta M_{\pi\pi}$=
$480.1\pm2.4\ $MeV, and a
width of $\sigma=20.9\pm2.6\ $ MeV. The upper signal has a yield
of $350^{+57}_{-55}$, $\Delta M_{\pi\pi}= 595.8\pm0.8\ $ MeV and $\sigma=4.2\pm0.7\ $ MeV.
All of these uncertainties are statistical, coming from the fit. 
The mass resolutions in these regions are $\approx 2.0$ and $\approx 2.8\ $MeV, 
respectively, based on our Monte Carlo simulation.
The lower peak clearly has a width greater than the experimental resolution. 
If we fit it
to a Breit-Wigner function, we obtain a width, $\Gamma$, of $\approx$ 50 MeV, 
but it can equally 
well be fit to a sum of more than one wide peak. 
If we fit the upper peak to a Breit-Wigner
convolved with a double Gaussian detector resolution function, 
we obtain a width of 
$\Gamma=4\pm2\pm2\ $ MeV where the uncertainties 
are statistical and systematic, 
respectively. The dominant
systematic uncertainty comes from uncertainties in the detector resolution 
function. This experimental width 
is not significantly different from zero; we place an upper limit
of $\Gamma < 8\ $ MeV at 90\% confidence level.
We estimate the 
systematic uncertainty on the
mass difference measurement of the upper state to be $\pm2\ $ MeV, due principally to
uncertainties in the 
momenta measurements and differences in the mass obtained using different fitting
procedures.

To help identify these new states, we investigate whether the decays 
proceed via intermediate $\Sigma_c$ and/or $\Sigma_c^*$ baryons. 
There is very little isospin 
splitting in the masses of these intermediate states, and, 
by isospin conservation, we 
expect equally many decays to proceed via a doubly charged $\Sigma_c^{(*)}$ 
as via a neutral one. 
To search for resonant substructure in the upper, narrower, state we 
use a signal mass band of
589$<$$\Delta M_{\pi\pi} $$<$ 603 MeV and sidebands of 
527$<$$\Delta M_{\pi\pi} $$<$ 575 MeV and
617$<$$\Delta M_{\pi\pi} $$<$ 665 MeV. 
This signal band has a signal yield of 314$\pm$50 .
We then plot the single $\pi$ mass difference, 
$\Delta M_{\pi} = M(\Lambda_{c}^{+}\pi^{\pm}) - M(\Lambda_{c}^{+})$
for both transition pions in the signal region and subtract the
sideband data, appropriately scaled.
The resultant plot (Figure 2) is fit to a sum of a polynomial background
and two signal shapes for the 
$\Sigma_c$
and $\Sigma_c^{*}$ baryons, with these shapes obtained by fitting
the inclusive $\Delta M_{\pi}$ plot, {\it i.e}., without
any cut on $\Delta M_{\pi\pi}$. The signal yields obtained by the fit
are $96\pm18$ and $-34\pm28$ events respectively. This gives a fraction 
of this state proceeding 
via an intermediate $\Sigma_c$ of $(31\pm 6 \pm 3)\%$, 
and an upper limit on the fraction proceeding
through $\Sigma_c^*$ of $11\%$ at 90\% confidence level. 
The dominant contribution to the systematic uncertainty in the
$\Sigma_c$ fraction is from our fitting procedures.
We cannot perform the same 
analysis for the lower state
because the limited kinematics of the decays makes kinematic reflections 
in the $\Delta M_{\pi}$
mass difference plots that the subtraction procedure 
cannot remove. 

We also display the data by first making a requirement 
of 163 $<$$\Delta M_{\pi}$$<$171 MeV and then 
plotting the dipion mass difference $\Delta M_{\pi\pi}$ (see Figure 3(a)). 
This requirement includes 
most of the decays that proceed via a $\Sigma_c$, but excludes the majority that 
decay non-resonantly to 
$\Lambda_c^+\pi^+\pi^-$. Figure 3(a) is fit to a sum of the two signal 
peaks, using fixed signal shapes and masses that were
found from Figure 1, and a polynomial background shape. 
The yield for the two signals $262\pm45$ and $105\pm16$ respectively. 
This second yield agrees well 
with the expectation from Figure 2, and confirms that a large fraction 
of the upper peak decays via $\Sigma_c\pi$. 
The yield of the lower peak also indicates
that it also resonates through $\Sigma_c$. 
We can also make a similar plot, using a cut on the single pion mass difference consistent
with being due to a $\Sigma_c^{*}$, 
namely 223 $<$$\Delta M_{\pi}$$<$ 243 MeV.  This is more 
problematical, because this mass window will include much of the phase-space available for
non-resonant decays, and will also not include the entire 
broad $\Sigma_c^*$ region. The dipion mass 
difference plot (Figure 3(b)), shows very little evidence of the upper peak, 
confirming the conclusion obtained from Figure 2. 
It does show considerable excess (331$\pm$47)
events in the region of the lower
peak, but it is difficult to calculate how much of this is really due to 
$\Sigma_c^*$. We display Figure 3 starting from $\Delta M_{\pi}=420\ $MeV to avoid 
irrelevant enhancements due to $\Sigma_c $ production that appears below
this threshold.
    
In summary, we find the lower peak to decay
resonantly via $\Sigma_c$ and probably also via $\Sigma_c^*$;
we cannot rule out 
a contribution from non-resonant $\Lambda_c^+\pi^+\pi^-$.  
The upper peak is comparatively narrow, and appears
to decay via $\Sigma_c\pi$ and to non-resonant
$\Lambda_c^+\pi^+\pi^-$, but not via $\Sigma_c^*\pi$.

Most models of charmed baryon spectroscopy start from the assumption 
that the baryon consists
of a heavy charm quark, and a light diquark which is itself in a well 
defined spin and parity
state, $J^P_{light}$. 
The decays that take place need to obey quantum mechanical decay rules 
for conservation of 
both $J^P$ and $J^P_{light}$ separately.
The lowest lying orbital excitations 
in the $\Sigma_{c}$ baryons should, like 
those of the $\Lambda_{c}$ baryons, have the  
unit of orbital angular momentum between the diquark and the charm quark; this 
will give five isotriplets.  
At higher masses, there should be 
five $\Lambda_c^+$ particles and two 
isotriplets of $\Sigma_c$ particles with L=1 between the two 
light quarks. Here we will 
refer to this second generation of orbital excitations as $\Lambda_c^{\prime}$
and $\Sigma_c^{\prime}$ states. 
Many of the $\Sigma_c$, $\Sigma_c^{\prime}$ and $\Lambda_c^{\prime}$ 
particles with L = 1 will decay rapidly and have large intrinsic 
widths. Only one undiscovered state in the 
sequence has no allowed two-body decays to a lower mass charmed baryon, 
and that is the 
$\Lambda_{c0}^{+ \prime }$, which has $J^{P}$$=$${1\over{2}}^{-}$ and 
$J^P_{light}$$=$$0^{-}$. This is therefore a candidate for the upper 
peak that we have found. 
Conservation of $J^P_{light}$, as required by Heavy Quark Effective
Theory, would not allow this particle to decay via $\Sigma_c\pi$. 
However, there is another 
state (the $\Lambda_{c1}^{+\prime}$)
with same overall quantum numbers, but this time with $J^P_{light}=1^-$, 
which is
expected to be at a similar mass. 
As the two states have the same quantum numbers, they might 
mix, and as the latter state can decay via an $S$-wave to $\Sigma_c\pi$, 
this could explain the 
fraction of decays of our peak resonating in that manner. 
Identification of the lower, wider, state is also
open to interpretation. 
One possibility is that is consists of a pair of $\Sigma_{c1}^+$
particles, with overall $J^P$$=$${1\over{2}}^-$ 
and $J^P$$=$${3\over{2}}^-$. These particles might be
expected to be split in mass by around 30 MeV, 
and should have preferred decay mode of 
$\Sigma_c\pi$ and $\Sigma_c^*(\pi)$
respectively. Their widths have been predicted to be around 100 MeV\cite{PIRJ}. 
We stress that there may be many other interpretations of our data, including the 
decay of radial excitations of charmed baryons.

In conclusion, we report the observation of structure in the 
$M(\Lambda_c^+\pi^+\pi^-)-M(\Lambda_c^+)$
mass difference plot, which we believe corresponds to the discovery of new 
excited charmed baryons. One enhancement, 
at $\Delta M_{\pi\pi} \approx 480\ $ MeV, is very wide ($\Gamma \approx 50\ $ MeV)
and it appears to resonate through $\Sigma_c$ and probably also $\Sigma_c^*$. 
The other, with a mass of $596\pm1\pm2\ $ MeV above the $\Lambda_c^+$, 
is much narrower ($\Gamma < 8$ 
MeV at 90\% confidence level), 
and appears to decay both via $\Sigma_c\pi$ and non resonantly to 
$\Lambda_c^+\pi^+\pi^-$, but not via $\Sigma_c^*$. 
We have no measurements of the spin and parity of these new states, but
we make educated guesses as to their identities.

\begin{figure}[htb]
\noindent
\psfig{bbllx=30pt,bblly=200pt,bburx=440pt,bbury=770pt,
file=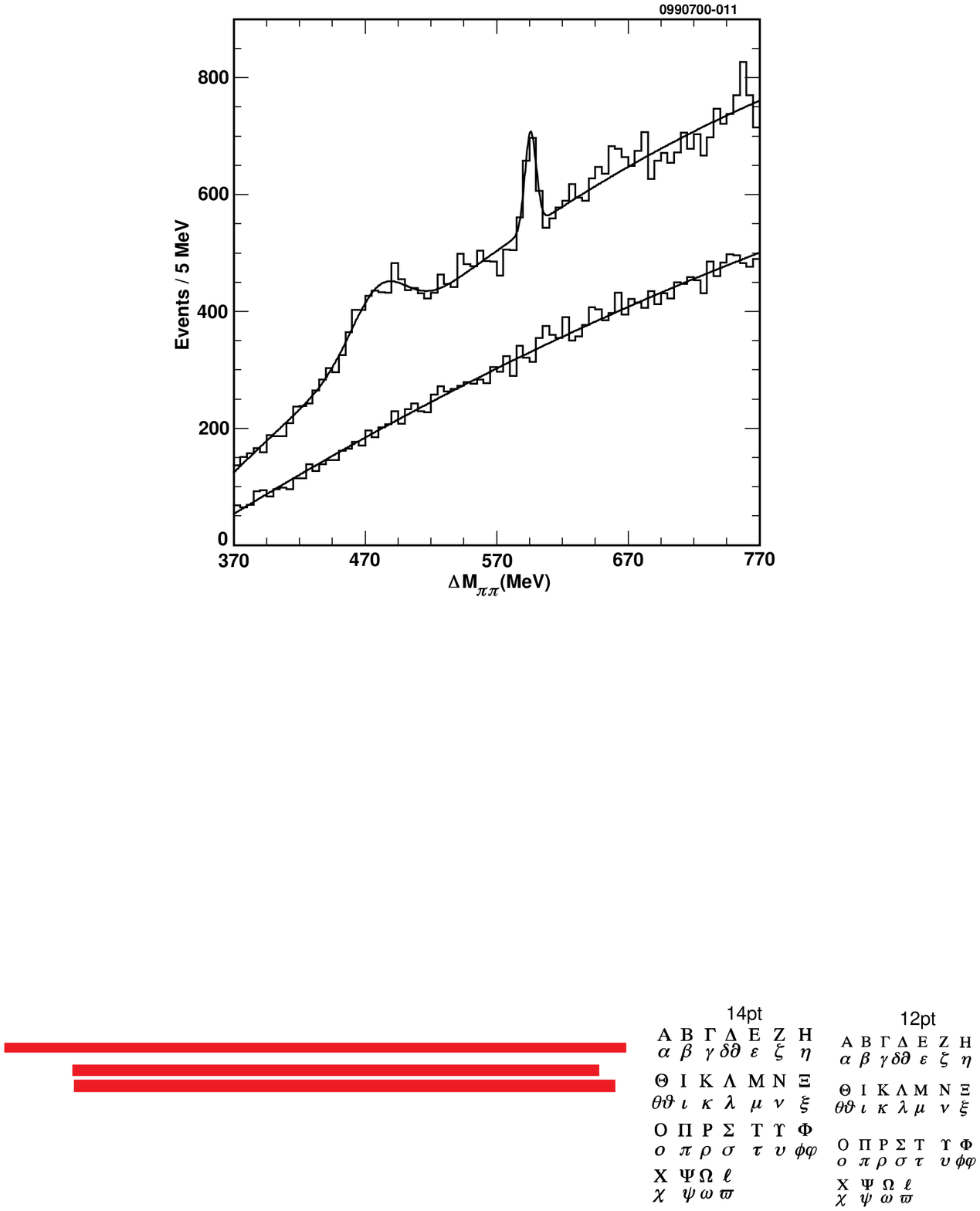,width=5.0in}
\caption[]{The upper histogram shows
$\Delta M_{\pi\pi} = 
M(\Lambda_{c}^{+}\pi^{+}\pi^{-})-M(\Lambda_c^+)$ 
above the $\Lambda_{c1}$ range;
the fit is to a quadratic background shape plus two Gaussian signal
functions. The lower histogram shows 
the same distribution for scaled $\Lambda_c^+$ sidebands, fit to a quadratic
background shape.}
\end{figure}

\begin{figure}[htb]
\psfig{bbllx=20pt,bblly=300pt,bburx=440pt,bbury=770pt,
file=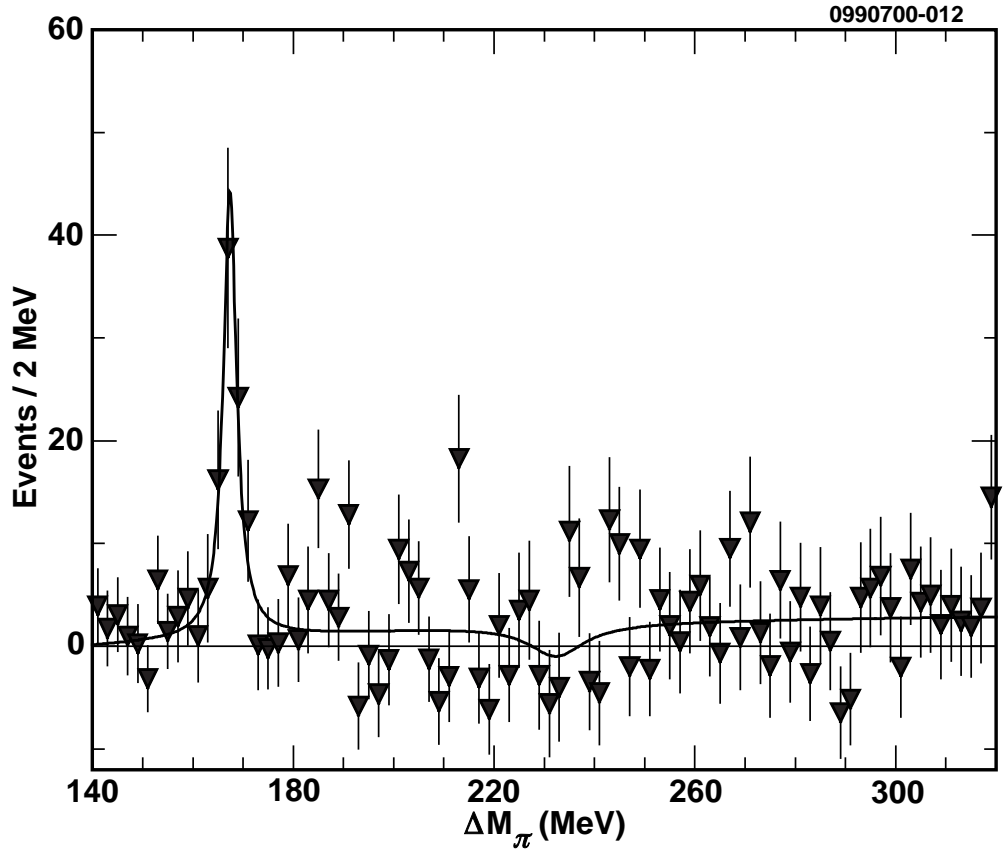,width=5.0in}
\caption[]{$\Delta M_{\pi}=
M(\Lambda_c^+\pi^{\pm})-M(\Lambda_c^+)$ in the upper resonance region, 
after sideband subtraction. }
\end{figure}

\begin{figure}[htb]
\psfig{bbllx=30pt,bblly=100pt,bburx=500pt,bbury=700pt,
file=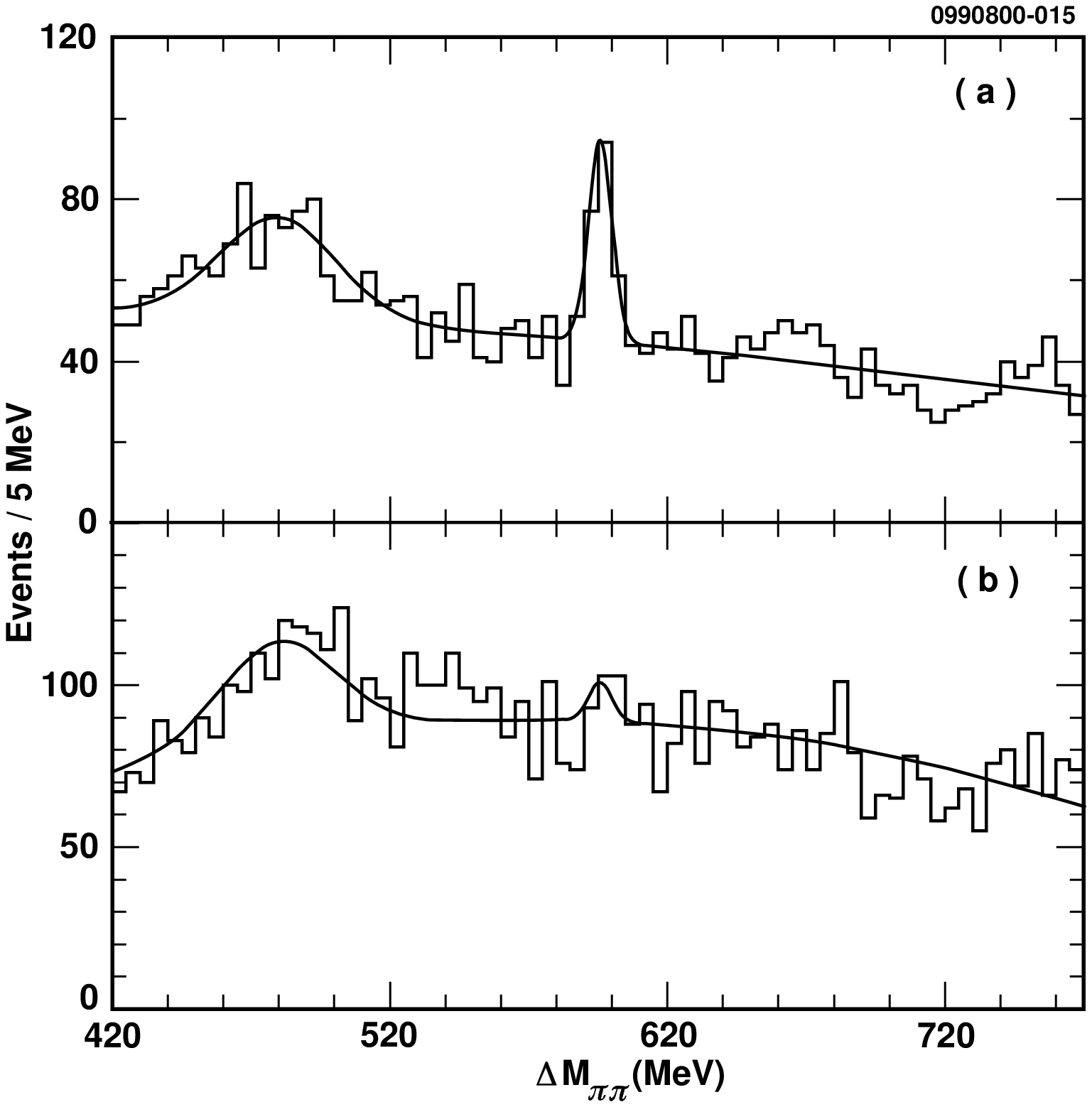,width=5.0in}
\caption[]{$\Delta M_{\pi\pi}=
M(\Lambda_c^+\pi^+\pi^-)-M(\Lambda_c^+)$ with cuts that (a) 
$\Delta M_{\pi} =  M(\Lambda_c^+\pi^)-M(\Lambda_c^+)$ 
is consistent with that expected for a $\Sigma_c$, and
(b)

$\Delta M_{\pi} = M(\Lambda_c^+\pi)-M(\Lambda_c^+)$
is consistent with that expected for a $\Sigma_c^*$.}
\end{figure}

\bigskip

We gratefully acknowledge the effort of the CESR staff in providing us with
excellent luminosity and running conditions.
This work was supported by 
the National Science Foundation,
the U.S. Department of Energy,
the Research Corporation,
the Natural Sciences and Engineering Research Council of Canada, 
the A.P. Sloan Foundation, 
the Swiss National Science Foundation, 
the Texas Advanced Research Program,
and the Alexander von Humboldt Stiftung.


\vfill

\begin{references}

\bibitem{LCS} 
H. Albrecht {\it et al.,} Phys. Lett. {\bf B}317, 227 (1993);
P. Frabetti {\it et al.}, Phys. Rev. Lett {\bf 72}, 961 (1994);
K.~Edwards  {\it et al.}, Phys. Rev. Lett {\bf 74}, 3331 (1995);
H. Albrecht {\it et al.,} Phys. Lett. {\bf B} 402, 207 (1997).
\bibitem{KUB} Y.~Kubota {\it et al.}, Nucl. Instrum. and Meth. A {\bf 320}, 66 (1992).

\bibitem{HILL} T.~Hill {\it et al.}, Nucl. Instrum. and Meth. A {\bf 418}, 32 (1998).

\bibitem{LAMC} P.~Avery {\it et al.}, Phys. Rev. D {\bf 43}, 3599 (1991); 
P.~Avery {\it et al.}, Phys. Rev. Lett. {\bf 71}, 2391 (1993); 
P.~Avery {\it et al.}, Phys. Lett. B {\bf 235}, 257 (1994);
M.S.~Alam {\it et al.}, Phys. Rev. D {\bf 57}, 4467 (1998).

\bibitem{GEANT} R. Brun {\it et al.}, GEANT 3.15, CERN Report No. DD/EE/84-1 (1987).

\bibitem{sigmastarplus}
R.~Ammar {\it et al.}, CLNS 00-1681, hep-ex/0007041,
submitted to Phys. Rev. Lett.

\bibitem{PIRJ} D.~Pirjol and T.-M.~Yan, Phys. Rev. D {\bf 56}, 5483 (1997).

\end{references}
\end{document}